\documentclass[preprint,showkeys,...]{revtex4-1}
\usepackage{graphicx}
\draft 

\begin{document}


\title{\large {\bf Realization of epitaxial ZnO layers on GaP(111) substrates by pulsed laser deposition}} 



\author {S. D. Singh\footnote{Corresponding author email: devsh@rrcat.gov.in}, R. S. Ajimsha, C. Mukherjee, Ravi Kumar, L. M. Kukreja, and Tapas Ganguli}
\affiliation{Raja Ramanna Centre for Advanced Technology, Indore, Madhya Pradesh-452013, India}


\date{\today}

\begin{abstract}

Epitaxy of ZnO layers on cubic GaP(111) substrates has been demonstrated using pulsed laser deposition. Out of plane and in-plane epitaxial relationship of ZnO layer with respect to GaP substrate determined using phi ($\phi$) scans in high resolution X-ray diffraction measurements are (0001)$_{ZnO}$ $\parallel$ (111)$_{GaP}$ and $\left\langle-12-10\right\rangle$$_{ZnO}$ $\parallel$ $\left\langle-110\right\rangle$$_{GaP}$ respectively. Our results of epitaxy of ZnO and its intense excitonic photoluminescence with very weak defect luminescence suggest that (111) oriented GaP can be a potential buffer layer choice for the integration of ZnO based optoelectronic devices on Si(111) substrates.      
\end{abstract}


\keywords{Epitaxy, Pulsed Laser Deposition, X-ray Diffraction, Wide band gap semiconductors}

\maketitle 


\section {Introduction}
 
For effective use of a material in optoelectronic devices, its crystalline and optical qualities are of paramount importance. These are achieved by growing epitaxial thin films of the material on appropriate substrates. Wurtzite ZnO is a wide band gap semiconductor with large exciton binding energy ($\sim$60meV)~\cite{siox, znosic} and it has been traditionally grown on hexagonal symmetry based substrates like ScAlMgO$_{4}$~\cite{ohtomo}, Al$_{2}$O$_{3}$~\cite{kukreja} and SiC~\cite{znosic}. Apart from this, (111) oriented cubic substrates~\cite{znogaas, znocubic} possessing three fold symmetry are also used for the epitaxial growth of ZnO, in which silicon (Si) is most widely used substrate~\cite{silit, silit1}. However, it is the presence of amorphous SiO$_{x}$ layer formed during the growth of ZnO on Si substrate, that results in completely random orientation of ZnO on Si substrate~\cite{silit, silit1}. In addition to this, large lattice mismatch ($\sim$15.4$\%$) and large diffenece in the thermal expansion coefficient ($\sim$48$\%$) between Si(111) and ZnO also pose another difficulty in realizing good quality ZnO layers~\cite{silit, silit1}. In view of this, several buffer layers of different materials like GaN~\cite{znogan}, AlN~\cite{znoaln}, Al$_{2}$O$_{3}$~\cite{znoal2o3}, ZnS~\cite{znozns}, MgO~\cite{znomgo}, Y$_{2}$O$_{3}$~\cite{silit}, CeO$_{2}$~\cite{znoceo}, Lu$_{2}$O$_{3}$~\cite{silit1}, Sc$_{2}$O$_{3}$~\cite{znosco}, YSZ~\cite{znoysz}, Gd$_{2}$O$_{3}$~\cite{znogdo} etc. have been explored to achieve epitaxial ZnO layer on Si(111) substrate. It is important to note that good epitaxial quality GaP layers have been grown on Si substrates, because of very small ($\sim$0.37$\%$) lattice mismatch between Si and GaP semiconductors.~\cite{sigap} Thus, GaP material can be a good choice as a buffer layer on Si substrates for the epitaxial growth of ZnO over it. For this, firstly the possibility of epitaxial ZnO on GaP substrate itself has been explored using pulsed laser deposition (PLD), which can have following advantages. 
There is a lattice mismatch of $\sim$15.8$\%$ between ZnO and GaP(111), which is comparable to that of $\sim$15.4$\%$ between ZnO and Si(111). Apart from this, there is a smaller difference ($\sim$7$\%$) between the thermal expansion coefficient of GaP and ZnO as compared to that of Si and ZnO. The thermal expansion coefficient of Si~\cite{thermalexpsigap}, GaP~\cite{thermalexpsigap}, and ZnO~\cite{thermalexpzno} are 2.6x10$^{-6}$/$^{0}$C, 4.65x10$^{-6}$/$^{0}$C, and 5.0x10$^{-6}$/$^{0}$C respectively. Additionally, the other inherent advantage of ZnO growth on GaP substrate may be that phosphorous (P) might diffuse from substrate into ZnO layer by post growth annealing process and could lead to p-type conductivity in the ZnO layer. Phosphorous doping has already been shown to produce p-type ZnO layers.~\cite{pdopingzno} Similar post growth annealing process of ZnO layers grown on GaAs substrate have produced p-type ZnO layers due to the out diffusion of arsenic from GaAs substrate into ZnO layer.~\cite{znodoping}
Hence, GaP substrate has enough technological advantage for exploring the growth of ZnO on it. Following this, we have recently reported a large value of valence band offset (2.81$\pm$0.2 eV) between ZnO and GaP hetero-junction using synchrotron based photoemission technique~\cite{sdsaplgap}. In this letter, we report the epitaxial growth of ZnO on GaP(111) oriented substrate using PLD and determine the epitaxial relationship of ZnO layer with respect to GaP substrate by performing high resolution X-ray diffraction (HRXRD) measurements. Thus, deposition temperature was varied in a range (300$^{o}$C - 500$^{o}$C) where a reasonable epitaxial quality of ZnO was observed.

\section {Experimental}

ZnO layers were grown on GaP(111) substrates by PLD. The PLD chamber was evacuated to a base pressure of $\sim$5$\times$10$^{-6}$ mbar using turbo molecular pump. The KrF excimer laser ($\lambda$=248 nm) beam of energy density $\sim$2 Jcm$^{-2}$ was focused on an in-house prepared high purity (99.9995$\%$) ceramic ZnO target for ablation. The substrate to target distance was kept at $\sim$4 cm and the target was rotated continuously during the deposition for uniform ablation. The growth temperature was varied from 300$^{o}$C - 500$^{o}$C to achieve a reasonable crystalline quality of ZnO layers on GaP substrates. The initial $\sim$35 nm growth of ZnO layer on GaP substrate was carried out at the desired substrate temperature in vacuum, to minimize the possible oxidation of GaP surface during the deposition of ZnO. This was followed by the growth of $\sim$170 nm thick ZnO over layer grown at the same temperature but in presence of oxygen at a partial pressure of $\sim$1$\times$10$^{-4}$ mbar. Crystalline quality and epitaxy of the deposited ZnO layers were assessed from HRXRD measurements, which were performed using PANalytical X'Pert Pro MRD diffractometer, equipped with a Hybrid monochromator (PANalytical model Hybrid 4x), with a beam divergence of about 18 arcsec perpendicular to the scattering plane
for Cu K$_{\alpha1}$ x-rays ($\lambda$= 1.5406 $\r{A}$). Surface morphology of the ZnO layers was investigated by using the atomic force microscopy (AFM) technique. AFM measurements were
carried out using a multimode scanning probe microscope (NTMDT, SOLVER-PRO). Silicon cantilever tips having a radius of curvature of $\sim$20 nm, resonant frequency $\ge$190 KHz and spring constant $\sim$5.5 Nm$^{-1}$ were used in a non-contact mode for the AFM measurement. Room temperature photoluminescence (PL) measurements were excited by using He-Cd laser ($\lambda$=325 nm) and detected by a CCD.

\section {Results and discussions}

Figure 1(a) shows 2theta/omega (2$\theta$/$\omega$) scan of ZnO layers grown at different temperatures on GaP(111) substrate along with a bare GaP(111) substrate. In addition to (111) and (222) reflections of GaP substrate and a peak from the background (marked as star), only (0002) and (0004) reflections of ZnO layer are observed in the investigated 2$\theta$ range. It is noted that the intensity of ZnO reflections relative to GaP substrate increases with the deposition temperature of ZnO layer. The (0004) reflection of ZnO is quite weak for the deposition temperature of 300$^{o}$C and it becomes prominent for the deposition temperature of 500$^{o}$C. These observations suggests that there is an improvement in the crystalline quality with increase in the deposition temperature and deposited ZnO layer are highly c-axis oriented. Figure 1(b) depicts an open detector omega scan for (0002) symmetric reflection of ZnO for all the samples. The open detector omega scan width determined from the XRD measurements are $\sim$2.6$^{o}$, $\sim$1.9$^{o}$, and $\sim$1.0$^{o}$ for the ZnO layers deposited at 300$^{o}$C, 400$^{o}$C, and 500$^{o}$C, respectively. Thus, the open detector omega scan width decreases with increase in the deposition temperature. The obtained width of $\sim$1.0$^{o}$ for ZnO deposited on GaP(111) substrate is comparable to the widths of good quality ZnO layers grown on Si, GaAs, and Ge substrates.~\cite{siox, znoxrd, sdsbandoffset} Hence, 2$\theta$/$\omega$ and open detector $\omega$ scans indicate that the crystalline quality of the ZnO layer improves with the deposition temperature. There is no signature of (10-11) asymmetric reflection of ZnO for the layer deposited at 300$^{o}$C because of the poor crystalline quality of the ZnO layer. Open detector omega scan for (10-11) asymmetric reflection of ZnO layers deposited at 400$^{o}$C and 500$^{o}$C are shown in Fig. 1(c). The (10-11) reflection of ZnO is quite weak for the deposition temperature of 400$^{o}$C, while it is relatively strong for the deposition temperature of 500$^{o}$C. Observation of (10-11) asymmetric reflection in the omega scan indicates that the ZnO layer grown at 400$^{o}$C and 500$^{o}$C has definite in-plane orientation with epitaxial nature. To confirm the epitaxial growth and in-plane epitaxial relationship of ZnO on GaP(111) substrate, phi ($\phi$) scan for (10-11) asymmetric reflection of ZnO layer and (220) reflection of GaP substrate are presented in Fig. 2. There are no peaks in the $\phi$ scan of ZnO layer deposited at 300$^{o}$C, while six broad peaks separated by $\sim$60$^{o}$ are observed for the deposition temperature of 400$^{o}$C. On the other hand, we clearly observe six sharp peaks separated by 60$^{o}$ corresponding to ZnO (10-11) reflection at the deposition temperature of 500$^{o}$C, which confirms the epitaxial growth of ZnO on GaP(111) substrate~\cite{silit1, znoceo}. The in-plane arrangement of reciprocal lattices and direct lattices of cubic GaP and wurtzite ZnO are shown in Fig. 3. The (111) plane of GaP(111) substrate along with the atoms on (111) plane indicating the triangular symmetry are shown in Fig. 3(a). Atomic arrangements on (111) plane and mutually perpendicular directions are shown in Fig. 3(b) for cubic GaP. Out of plane shows the [111] direction, while three sets of mutually perpendicular directions are [-110], [-1-12]; [10-1], [-12-1]; [0-11], [2-1-1] as indicated in Fig. 3(b). The arrangements of reciprocal lattice points as determined by the $\phi$ scan (Fig. 2) is depicted in Fig. 3(c). Out of plane shows the [111] directions, while two other mutually perpendicular directions are chosen to be [-110] and [-1-12]. The $\phi$ scan of (220) reflection of GaP(111) substrate shows three peaks separated by 120$^{o}$. The three reciprocal lattice points corresponding to these peaks are identified to be (022), (202), (220). One can generate (022) reciprocal lattice point using three mutually perpendicular directions [111], [-110], and [-1-12] as (022)=1$\times$(-110)+ 1/3$\times$(-1-12) + 4/3$\times$(111). Similarly, (202)=-1$\times$(-110)+ 1/3$\times$(-1-12)+ 4/3$\times$(111) and (220)= -2/3$\times$(-1-12) + 4/3$\times$(111). The three reciprocal lattice points (022), (202), and (220) of GaP are shown by black filled circles in Fig. 3(c). The $\phi$ scan of (10-11) reflection of ZnO shows six peaks, which correspond to six reciprocal lattice points of (-1,1,0,1), (0,1,-1,1), (1,0,-1,1), (1,-1,0,1), (0,-1,1,1), and (-1,0,1,1) that are also shown by the red filled circles in Fig. 3(c). Based on the fact that the angular ($\phi$) positions of (10-11) reflection of ZnO and (220) reflection of GaP substrate coincide with each other as obvious from Fig. 2, the arrangement of reciprocal lattice points of ZnO and GaP is depicted in Fig. 3(c). Following the arrangements of reciprocal lattice points of ZnO and GaP substrate, the in-plane superimposition of direct lattices of ZnO and GaP are shown in Fig. 3(d). Different in-plane directions of hexagonal ZnO are also indicated. It is clear from Fig. 3(d) that [-12-10]$_{ZnO}$ and [-110]$_{GaP}$ directions are parallel to each other. The another way to find out the in-plane epitaxial relationship is that the zone axis associated with (0001) and (10-11) reflections of wurtzite ZnO is [-12-10]~\cite{silit1, znoceo}. On the other hand, the zone axis corresponding to (111) and (220) reflections of cubic GaP is [-110]~\cite{silit1, znoceo}. As evident from the $\phi$ scans shown in Fig. 2, the angular ($\phi$) positions of (10-11) and (220) reflections of ZnO and GaP respectively coincide with each other. This confirms that the two axises [-12-10]$_{ZnO}$ and [-110]$_{GaP}$ are parallel to each other. Thus, the in-plane epitaxial relationship of $\left\langle-12-10\right\rangle$$_{ZnO}$ $\parallel$ $\left\langle-110\right\rangle$$_{GaP}$ is determined based on the $\phi$ scans as presented in Fig. 2.    

Figure 4 shows room temperature PL spectrum of ZnO layer deposited at 500$^{o}$C as a representative. A strong near band edge (NBE) luminescence with very weak defect luminescence is observed at RT from the ZnO layer epitaxially deposited on GaP(111) substrate. The FWHM of room temperature PL spectrum is about 170 meV, which is comparable to the reported values of good quality ZnO hetero-epitaxial layers. Additionally, there is no significant variation of optical properties on the deposition temperature. Inset to the Fig. 4 shows the surface topography of the epitaxial ZnO layer deposited at 500$^{o}$C as a representative, where elongated grains of average sizes of 300$\pm$50nm and 110$\pm$10nm are observed.  It is to be noted that good crystalline quality of GaP epitaxial layers on Si substrates have already been achieved by several groups including us.~\cite{sigap,sigap1, sigap2} Good epitaxial quality of ZnO on GaP substrate demonstrated in this paper opens up a possibility of ZnO/GaP/Si combination, where advantages of both GaP and Si materials can be utilized.                      

\section {Conclusion}     
 
Epitaxial ZnO layers on GaP(111) cubic substrate have been realized by varying the deposition temperature using PLD. The out of plane and in-plane epitaxial relationship of ZnO with respect to GaP substrate determined using $\phi$ scan in HRXRD experiments were (0001)$_{ZnO}$ $\parallel$ (111)$_{GaP}$ and $\left\langle-12-10\right\rangle$$_{ZnO}$ $\parallel$ $\left\langle-110\right\rangle$$_{GaP}$ respectively. Intense NBE PL with very low defect luminescence at RT was observed from epitaxial quality ZnO film, which also confirmed the good epitaxial quality of ZnO.

\begin{acknowledgments}

The authors acknowledge Dr. T. K. Sharma, Dr. V. K. Dixit, and Shri Himanshu Srivastava for useful discussions. The authors thanks Shri Vikas Sahu for his help in the PL measurements. The authors also acknowledge Dr. P. D. Gupta, Director, RRCAT, Dr. S. M. Oak, and Dr. S. K. Deb for his constant support during the course of this work.  

\end{acknowledgments}

\newpage
\bibliographystyle{apsrev}

\begin{figure}
 \includegraphics[width=1.0\textwidth]{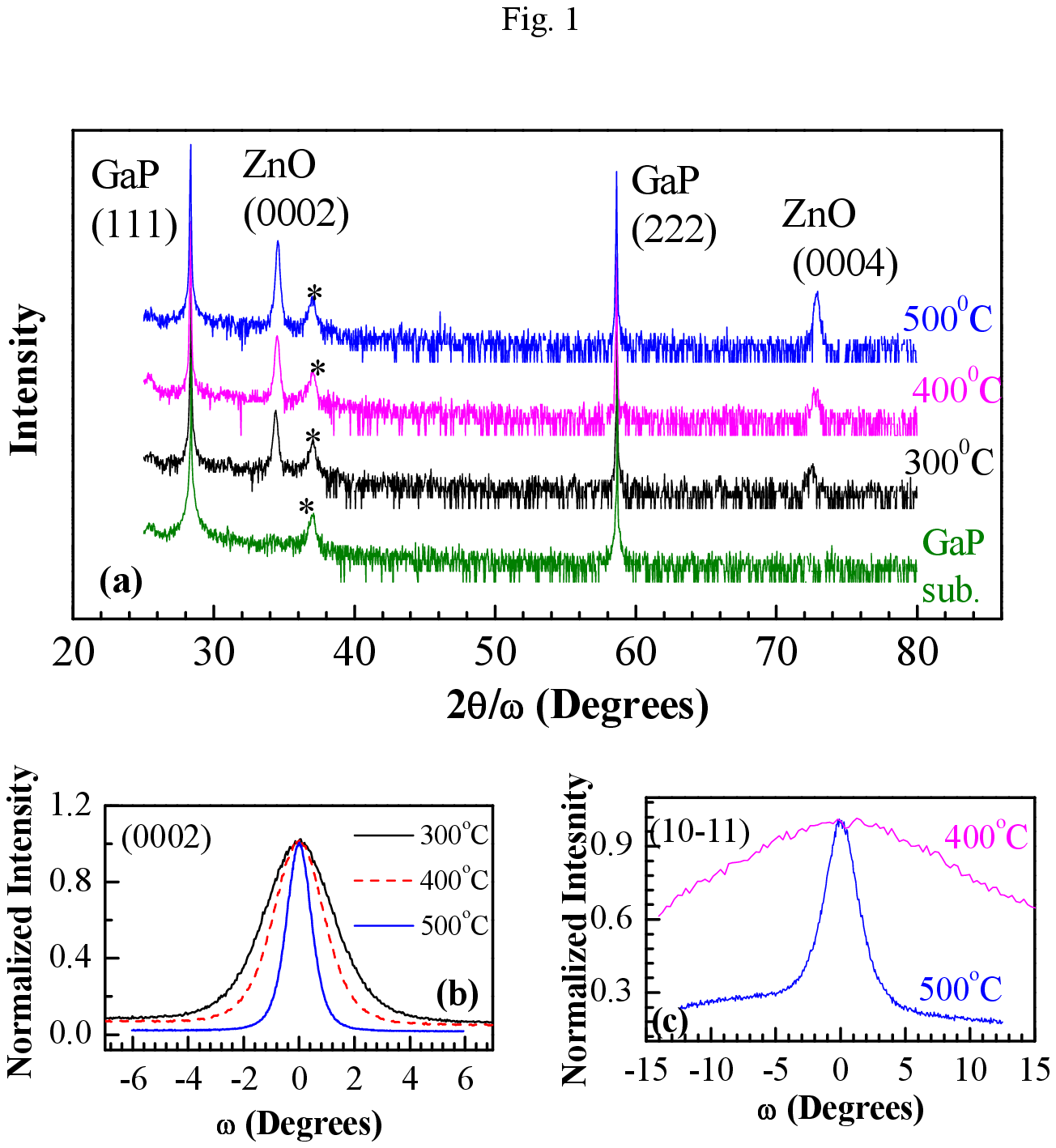}%
 \caption{(a) 2$\theta$/$\omega$ scan for ZnO layers deposition at different temperatures and bare GaP(111) substrate. A peak from the background marked as star is also shown. (b) Open detector $\omega$ scan for (0002) symmetric reflection of ZnO of layers deposited at different temperatures. Open detector $\omega$ scan for (10-11) asymmetric reflection of ZnO of layers deposited at different temperatures.}%
 \end{figure}

\newpage
\begin{figure}
 \includegraphics[width=1.0\textwidth]{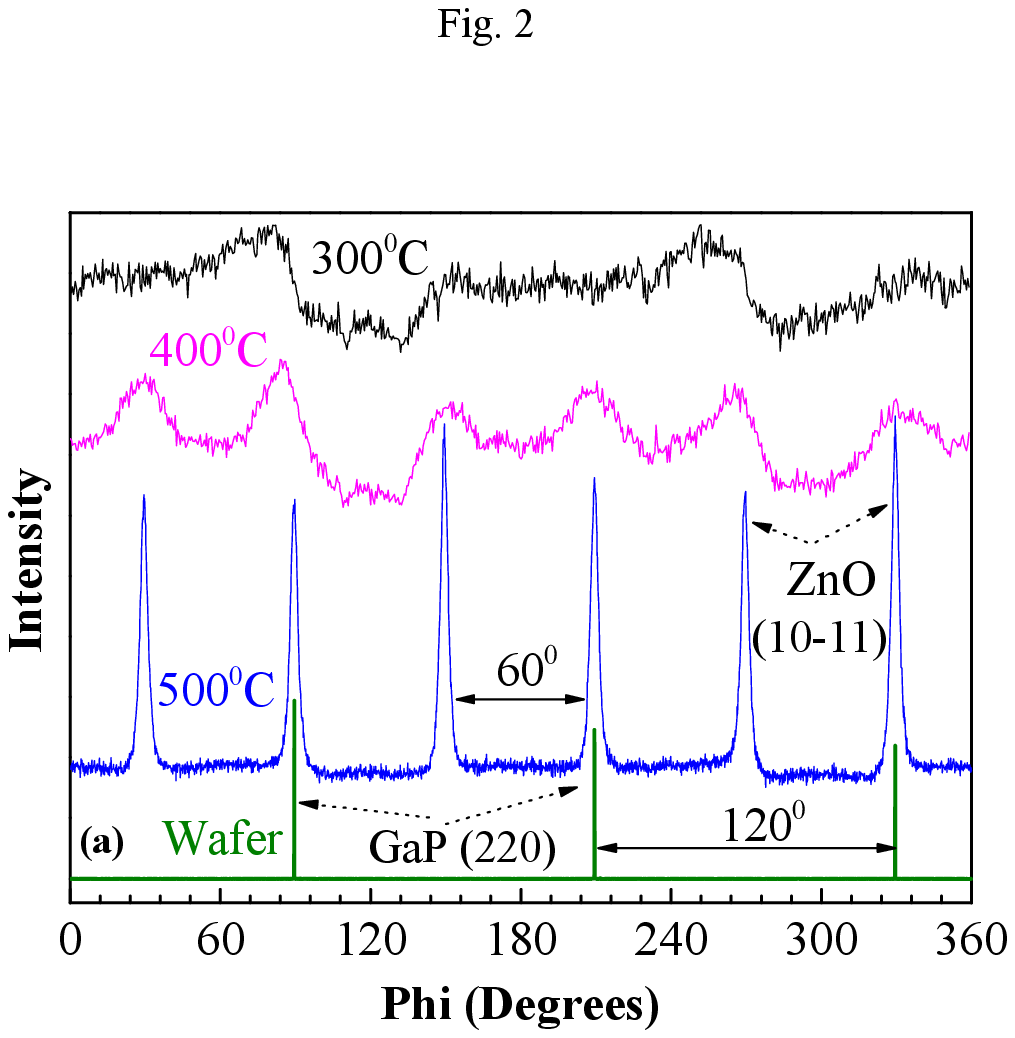}%
 \caption{Phi ($\phi$) scan of (10-11) asymmetric reflection of ZnO for layers deposited at 300$^{o}$C, 400$^{o}$C and 500$^{o}$C.}%
 \end{figure}
 
 \newpage
\begin{figure} 
 \includegraphics[width=1.0\textwidth]{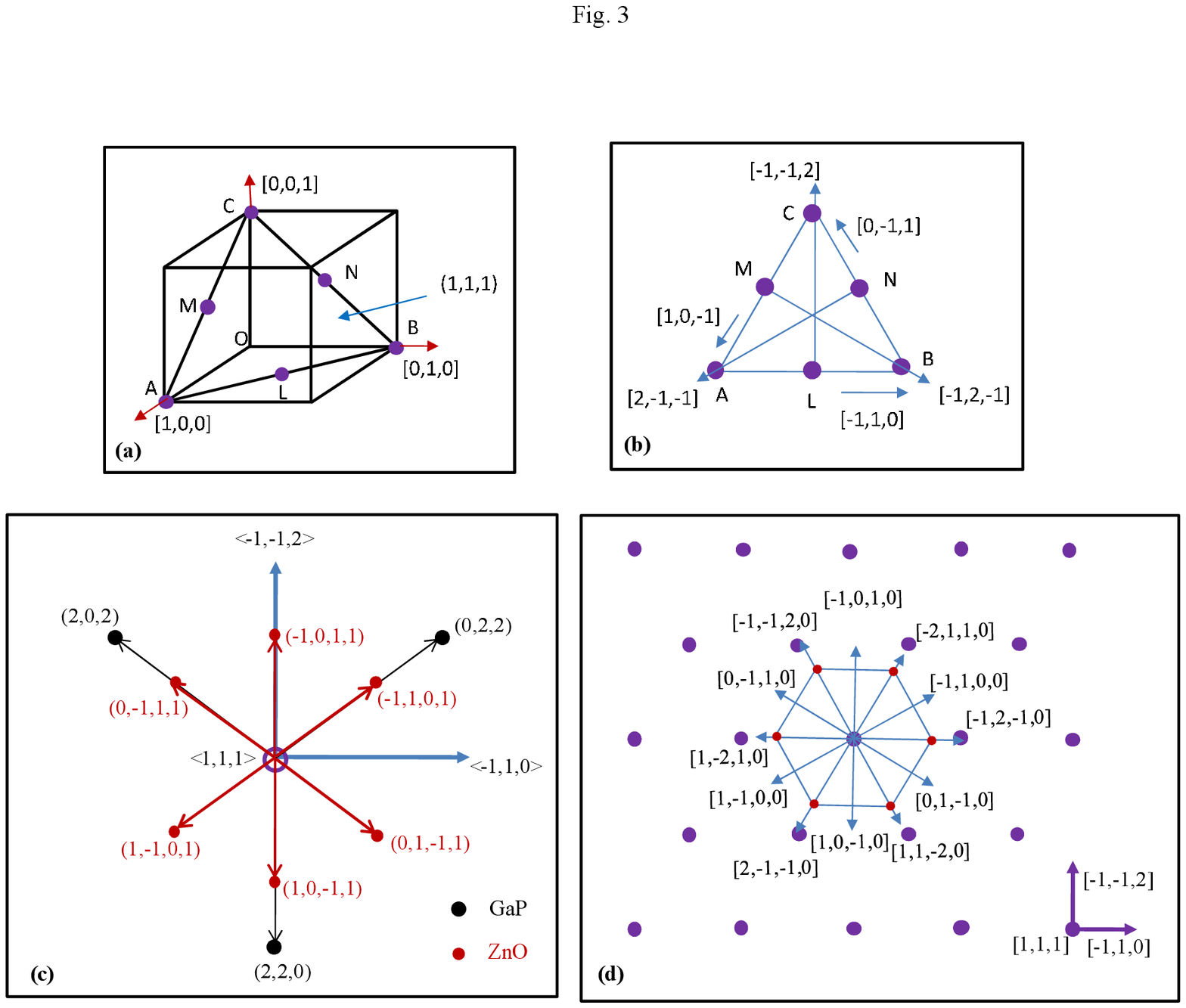}%
 \caption{(a) Cubic structure of GaP semiconductor showing (111) plane and atomic positions on this plane. [001], [101], and [100] directions are also marked. (b) Set of mutually perpendicular directions such as [-110], [-1-12]; [10-1], [-12-1]; [0-11], [2-1-1] are shown on the (111) plane of cubic GaP structure. Out of plane indicates [111] direction. (c) Arrangements of (10-11)$_{ZnO}$ and (220)$_{GaP}$ reciprocal lattice points. (022), (202), and (220) are the three reciprocal lattice points corresponding to three peaks in the phi($\phi$) scan of (220) reflection of GaP substrate. (-1,1,0,1), (0,1,-1,1), (1,0,-1,1), (1,-1,0,1), (0,-1,1,1), and (-1,0,1,1) are the six reciprocal lattice points corresponding to six peaks in the phi($\phi$) scan of (10-11) reflection of ZnO epitaxial layer. (d) In-plane atomic arrangements of cubic GaP and wurtzite ZnO. Different directions of wurtzite ZnO are marked on the hexagon and (0001) direction for ZnO is out of the plane. Directions for cubic GaP are also marked at the right bottom and (111) direction is out of the plane.}%
 \end{figure}
 
 \newpage
\begin{figure}
 \includegraphics[width=1.0\textwidth]{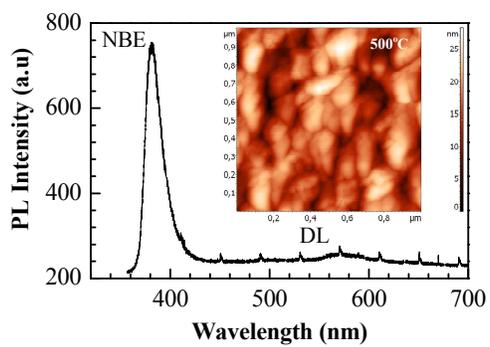}%
 \caption{Room temperature PL spectrum displaying intense NBE PL with very weak defect luminescence (DL) of epitaxial ZnO layer deposited at 500$^{o}$C. Inset shows the surface topography of the same epitaxial ZnO layer.}%
 \end{figure} 
 
%



\end{document}